\documentstyle[aps,prl,psfig]{revtex}

\newcommand{\be}{\begin{equation}}
\newcommand{\ee}{\end{equation}}

\newcommand{\bex}{\begin{eqnarray}}
\newcommand{\eex}{\end{eqnarray}}
\begin{document}

\title{Single-particle Nonlocality: A Proposed Experimental Test}
\author{R. Skrintha\thanks{e-mail: srik@iiap.ernet.in}}
\address{Indian Institute of Astrophysics, 
Koramangala, Bangalore- 34, Karnataka, India.}
\maketitle \date{}


\begin{abstract} 
We show that controlled interference of a particle's wavefunction can be used 
to perform a
quantum mechanical measurement in an incomplete basis. This happens because the
measurement projects the particle into a lower dimensional subspace of the 
Hilbert space of the incoming wave. It allows a sender (Alice) to signal the 
receiver (Bob) nonlocally, by Alice's choosing to measure in a complete or 
incomplete basis (in general: bases of differing incompleteness). If 
experimentally confirmed, it furnishes a new
quantum communication act: nonlocal transmission of a bit without
concomitant causal communication. However, the question of its compatibility
with special relativity remains unresolved.  
\end{abstract}

\parskip 2pt


\section{Introduction}
It has been known for a long time that two \cite{epr}
distant particles, when related by the property of quantum entanglement,
can display remarkable nonlocal behaviour, eg., the violation of Bell
type inequalities \cite{bel64}. Interestingly, a
formally similar phenomenon, in which the entangled twin is replaced by
the vacuum, has been predicted for single particles, too \cite{har94}.  In 
contrast, the single-particle nonlocality we talk about in the present article
does not involve any entanglement. Instead, it refers to the familiar 
sudden localization of 
a particle wavefunction following a position measurement. As a result, the
nonlocal effect we envisage is not of a two-party
correlation type, but involving 
probability modulation in a quantum informationally non-trivial way.
  
The article is arranged as follows.
The experimental set-up for and the basic idea behind
the test are presented in the next section. 
The main result is derived more rigorously in Section \ref{main},
and physically interpreted in Section \ref{degenr}, where we show that it owes 
its origin to measurement via in an incomplete basis via controlled 
interferometry. We then conclude with a brief final section.

\section{Experiment}\label{xper}
The proposed experimental test is quite simple. It
involves a light beam (preferably laser, in view of its low divergence),
of width $w$, from a spatially coherent source
being split up into three parallel beams ($a$, $b$ and $h$) by means of 
two 50/50 beams splitters (BS1 and BS2), as shown in Figure \ref{tri}. Beam
$a$ ($b$) is deflected by mirrors M0, MA45 and MA60 (MB45 and MB60) so that
both beams illuminate the horizontal region $k$.
Mirrors M0, MA45 and MB45 are inclined to the
horizontal at $\pm 45^{\circ}$, mirrors MA60 and MB60 at $\pm 60^{\circ}$. The
size of region $k$ is therefore $w^{\prime} \equiv w\sec 30^{\circ}$.

Beams $a$ and $b$ are observed by an observer, denoted Alice, and beam $h$
by another, denoted Bob. Alice is equipped with a movable vertically
oriented detector 
system that can be positioned either at the position DB2, which covers region
$k$, or DB1, which lies ahead
 (Figure \ref{tri}). Path length compensator PLC is provided on
beam $b$, so that both beams are in phase when incident on MA45 and MB45.
After reflection from mirror MA60 (MB60), beam $a$ ($b$) has a vertical
wavefront (at angle 120$^{\circ}$, rather than 90$^{\circ}$, to the direction
of propagation). In this way,
we can ensure that both beams $a$ and $b$ interfere constructively at {\em all}
points in region $k$. Careful orientation of the mirrors and positioning of
detector DA2 is crucial to the experiment.

The action of the beam splitters and the mirrors
leaves the input photon as it approaches the detectors in the superposition 
state
\be
\label{output}
|\psi\rangle = \int\left(\frac{\lambda^{\prime}}{2}(|a(z)\rangle + |b(z)\rangle)
+ \frac{\lambda}{\sqrt{2}}|h(z)\rangle\right)dz,
\ee           
where $|X(z)\rangle$ $(X = a, b, h)$ labels a ray state at height $z$ in 
beam $X$, and $\lambda$ and $\lambda^{\prime}$
are the linear amplitude densities of the beams.

Suppose Alice
positions her detector at position DB1. If she detects a photon, it will 
be either at region $l$ or $m$. Because of the property of the beam splitters,
the probability that Alice finds a photon in the arm $a$ or $b$, 
$P(a) = P(b) = 1/4$, whilst 
the probability $P(h)$ that Bob finds a photon in the arm $h$ is $1/2$.
Writing
\be
P(a) = P(b) = \int_{w^{\prime}} |\lambda^{\prime}/2|^2dz = 1/4;
\hspace{1.0cm}
P(h) = \int_w|\lambda/\sqrt{2}|^2dz = 1/2.
\ee
we find $\lambda = 1/\sqrt{w}$ and $\lambda^{\prime} =
1/\sqrt{w^{\prime}}$. The photon count rate in 
each arm is given by $C(a) = C(b) = (1/2)C(h) = C/4$ 
where $C$ is the photon count rate for the light entering BS1 from 
the source.

On the other hand, suppose she positions her detector at position DB2.
She detects clicks only at region $k$, but
she cannot distinguish whether her detection was produced by a 
photon coming through beam $a$ or $b$. By Feynman's dictum that quantum
mechanics (QM) requires 
that all indistinguishable transition processes interfere \cite{fey65},
the beams interfere at $k$. Note that the mirrors are so oriented  
as to ensure that both beams interfere constructively at
{\em all} points in $k$, assuming that the laser beams diverge insignificantly
across the experiment. (If they diverge considerably,
then path length compensators with thickness varying suitably
across the light stream are inserted to ensure that interference at $k$ is
in-phase at all points, which is always possible.) On account of interference, 
Alice expects to find a photon with probability 
\be 
\label{one}
P(k) \propto \int_{w^{\prime}}\left|\frac{\lambda^{\prime}}{2} +
\frac{\lambda^{\prime}}{2}\right|^2dz = 1.
\ee
At the same time, we still have
$P(h) \propto 1/2$.
We need to renormalize the probabilities because $P(k) + P(h) = 1 + (1/2) 
\equiv \beta = 3/2 \ne 1$. Thus, $P(k) = 1/\beta = 2/3$ and $P(h) = 1/(2\beta)
= 1/3$. Consequently, in this case, $C(k) = 2C/3$ and $C(h) = C/3$. 

It is somewhat surprising that we have to renormalize the probabilities. Yet,
later we shall see that the renormalization is inevitable on account
of the incompleteness of the DA2 measurement basis. 
It would not be needed if, instead of the DA2 measurement, Alice 
uses a Young double slit set-up, wherein beam $a$ passes
through one slit, and beam $b$ through another, to interfere on a screen lying
ahead. Note that this also, as with the DA2 measurement, entails that the
two beams are indistinguishably registered. But, as shown in the next section, 
the pattern of bright and dark fringes on the screen will be such that the
integrated count will be $C/2$, so that $C(h)$ also remains $C/2$.
One aspect of the difference between the two measurements is that the DA2 
measurement is a {\em controlled} interference, in which the relative phase 
of the interfering rays is pre-determined, whilst in the double slit 
interference, being diffraction determined, it is not.

The renormalization requires a subtle shift in the probabilistic interpretation
of the wavefunction. The conventional statement of probability conservation
in the region of spatial overlap of beams $a(x, t)$ and $b(x, t)$
\be
\label{conserv}
\frac{\partial}{\partial t}(a + b)(a^{*} + b^{*})
= \frac{i}{\hbar}\bigtriangledown\cdot\left\{
(a^{*} + b^{*})\bigtriangledown(a + b) -
(a + b)\bigtriangledown(a^{*} + b^{*})\right\}.
\ee
contains no reference to beam $c$. This raises the 
question how the $a$-$b$ interference can deplete probability in arm 
$c$. The answer, which is elucidated by the second quantization
formalism, is that amplitudes are in this case not squared probabilities,
but only {\em proportional} to squared probabilities. In general, they are
{\em first order correlation functions}. Thus, Eq. (\ref{conserv}) is, 
strictly speaking, not the equation for probability conservation, but that for
first order correlation. It gives us a measure of how correlated a photon
being found at $k$ is with the input photon.
In the above experiment, the enhanced correlation in $k$ is compensated for
by conservation-enforcing correlation current flows into $k$ according to Eq.
(\ref{conserv}). However, correlation density at $p$ remains unchanged,
leading effectively to probability flow from the lower arm to upper arms.

In a double slit experiment, the pattern of bright-dark fringes indicate
phase difference between the interfering modes. These phase difference terms,
when integrated over, vanish, as expected of orthogonal states in Hilbert
space. Hence it is worth pointing out that the fact that
only constructive interference takes place at all points in $k$
does not violate the Hilbert space orthogonality
of the beams. If $X(x, t)$ is the wavefunction of beam $X$ ($X = a, b, h$), 
then unitarity guarantees orthogonality, i.e., that the volume integral
over all configuration space
\be
\label{overlap}
\int a(x, t)b^{*}(x, t)d^3x = 0.
\ee
Thus, even though the two beams interfere constructively in a small volume
surrounding region $k$, at some other regions of their overlap away from
the screen they interfere
destructively. This pattern of fringes in space will be such that
Eq. (\ref{overlap}) is non-trivially satisfied. 

It is interesting that by merely splitting and indistinguishably
re-combining her beam, Alice 
boosts her probability of finding a photon from $1/2$ to $2/3$, while
weakening Bob's from 1/2 to 1/3 without directly acting on his beam. 
But this also means that Bob can instantaneously 
discern whether Alice measured at DB1 or DB2,
depending on whether he finds count rate $C/2$ or $C/3$ on arm $h$.
Here we note that this nonlocal effect does not affect the strictly local
character of unitary evolution. It is during the DA2 measurement and the
consequent renormalization that a nonlocal connection between beams $a$, $b$ on 
the one hand, and $c$ on the other becomes manifest. 
It is as if the interferometer beams must
somehow talk to each other, no matter how far apart they are,
so that probabilities remain consistent with the correlation functions.

Suppose, on grounds of Einstein causality, we require that Bob's observed
count rate on arm $h$ should remain $C/2$. Then, because
$P(k) = 2\times P(h)$, Alice should observe a count rate $C$
at $k$. The combined intensity output for Alice and Bob would then be
$3C/2$, which is
greater than the input rate, in contradiction with energy conservation.
On the other hand, what about somehow prohibiting the interference of the 
re-combining beams? This option is unlikely because the recombining beams
are indistinguishably registered. It is not clear that we can prohibit
interference at $k$ without also prohibiting the familiar double slit
interference, where indistinguishability between the slits guarantees 
interference, in agreement with Feynman's above-mentioned dictum, and also
consistent with the quantum optical formalism, as shown in the next Section.

Finally, we note that if we insert a phase shifter of phase
$\phi$ in path $a$ at position DA1 in conjunction with detector DA2, Bob's
observed count rate is $C(2 + \cos\phi)^{-1}$. In this way,
the above experiment can be generalized for the transmission of 
continuum, as against discrete, signals. 


\section{Derivation}\label{main}

\begin{figure}
\centerline{\psfig{file=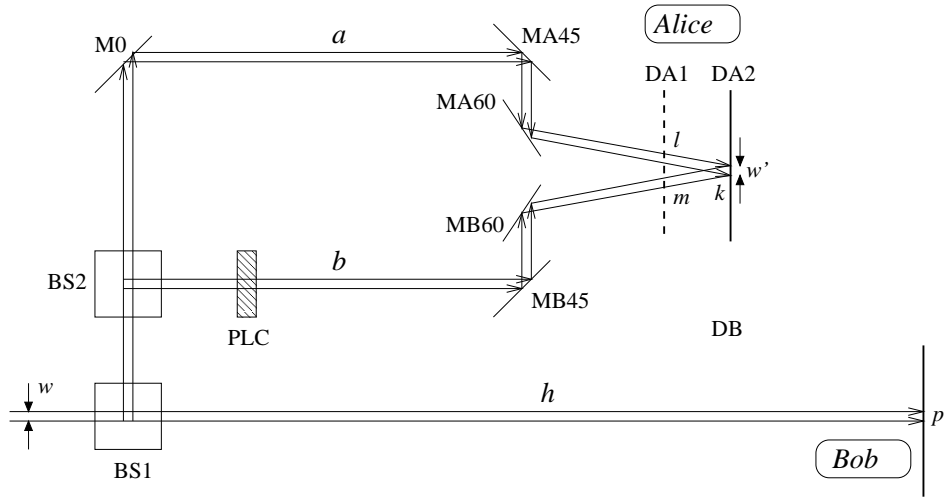,width=12.5cm}}
\vspace*{0.5cm}
\caption{A beam of light of width $w$, entering from the left,
 is split into two at beam splitter BS1. The
outgoing upper beam is in turn split into two beams, $a$ and $b$
at beam splitter BS2. Alice (Bob) observes $a$ and $b$ ($h$). 
The mirrors M0, MA45 (MB45) and MA60 (MB60) are used to fold
ray $a$ ($b$), and get them to converge to region $k$.
Alice can choose to detect her two beams either indistinguishably at $k$, by 
positioning her detector at DB2, or at regions $l$ and $m$, by positioning it at
DB1. Path length compensator PLC on 
path $b$ and careful orientation of the mirrors ensure that beams $a$ and $b$ 
illuminate region $k$ in-phase. Bob detects beam $h$ using a detector
positioned at region $p$. Mirrors M0, MA45 and MB45 are tilted at 45$^{\circ}$
to the horizontal, MA60 and MB60, at 60$^{\circ}$. Hence region $k$, assuming
that the light beams spread very little during their transit, has a
vertical extent $w^{\prime} = w\sec 30^{\circ}$.}
\label{tri}
\end{figure}

We now derive more rigorously the result
arrived at in the preceding section. The three-mode quantum optical 
state vector of the photon field just after leaving the beam-splitters
is given by:
\be
\label{spdc}
|\Psi\rangle = |{\rm vac}\rangle + \epsilon(|a\rangle + |b\rangle
+ \sqrt{2}|h\rangle ),
\ee
where $|{\rm vac}\rangle$ is the vacuum state, $|X\rangle$ ($X = a, b, h$)
is the Fock state mode corresponding to  some ray $z$ in beam $X$ of the 
interferometer and $\epsilon$ depends on the source strength \cite{gla63}.
Let us consider configuration \#1, wherein Alice positions her detector at DA1.
The positive frequency part of the electric field at any of the three detectors
is:
\be
\label{bobfjeld}
E_Y^{(+)} = \lambda^{\prime}e^{ikr_Y}\hat{Y}~~(Y = a, b); 
\hspace{1.0cm}
E_h^{(+)} = \lambda e^{ikr_h}\hat{h}.
\ee
where $\hat{X}$ is the annihilation operator for the $|X\rangle$ mode and
$r_X$ is the distance from the source to the detector (located at $l$, $m$
or $p$) along arm $X$ to a point $z$ on the detector (Figure \ref{tri}).

The single particle detection rate at $z$, or intensity
 $I(X)$ $(X = a, b, h)$, obtained by Alice or Bob, 
is proportional to the first order correlation function,
i.e, $\langle E^{(-)}_XE^{(+)}_X\rangle$, where $\langle\cdots\rangle$ indicates
the expectation value in the state $|\Psi\rangle$. 
The total photon count rate $C(X)$ is obtained
by integrating the intensity over the width of beam $X$:
\be
\label{Rg}
C(Y) \propto \int_{w^{\prime}}
\left\langle E_Y^{(-)}E_Y^{(+)}\right\rangle dz = \epsilon^2,\hspace{0.5cm}
C(h) \propto \int_w \left\langle E_h^{(-)}E_h^{(+)}\right\rangle dz = 
2\epsilon^2 ,
\ee
where we have used Eqs. (\ref{spdc}) and (\ref{bobfjeld}).
Thus, the chances of detection on any one of the arms $a$ and
 $b$ is half that on arm $h$.
On the other hand, let us consider configuration \#2, wherein 
Alice positions her detector at DA2. In this case, 
the positive part of the electric field at $k$ is given by:
\be
\label{fjeld2}
E_k^{(+)} = \lambda^{\prime}\left(e^{ikr_{ak}}\hat{a} + 
e^{ikr_{bk}}\hat{b}\right),
\ee
where $r_{Yk}$ ($Y = a, b$) is the distance from the source to region $k$ along
a ray $z$ in beam $X$, while Bob's field remains as in Eq. (\ref{bobfjeld}). 
The single counting rate at the region $k$ is:
\be
\label{Rf}
C(k) \propto \int_w \left\langle E_k^{(-)}E_k^{(+)}\right\rangle dz = 
    4\epsilon^2,
\ee
assuming that, by the appropriate setting of path length compensator PLC 
in Figure \ref{tri}, $r_a = r_b + nk^{-1}$, for some integer $n$.
The expression for arm $h$ remains as in Eq. (\ref{Rg}).
In an actual implementation, Eqs. (\ref{Rg}) and 
(\ref{Rf}) must be further modified to take
into consideration the single slit diffraction effects and the profile of the
laser beam.

Comparing Eqs. (\ref{Rg}) and (\ref{Rf}), we find that by measuring at $k$,
Alice has a two-fold greater chance of finding a photon than does Bob at $h$.
The total correlation is $\beta \equiv
 6\epsilon^2$. Since the total power has to be
 conserved, Alice's controlled interference boosts her observed rate to 
$4\epsilon^2C/\beta = 2C/3$, thereby
producing a corresponding $2/3$ factor depletion in Bob's photon counts.
We thus confirm the result of the preceding section. Similarly, we can also
verify the result for the phase shifter being inserted into path $a$ at DA1.

Suppose we replace DA2 with a Young double slit system, wherein $a$ passes
through one slit, and $b$ through another. The electric field, $E_y^{(+)}$,
at some point $y$ on the screen of size $S$, would be
\be
\label{fjeld3}
E_y^{(+)} = \frac{1}{\sqrt{S}}\left(e^{ik(r_{aa} + r_{ay})}\hat{a} + 
e^{ik(r_{bb} + r_{by})}\hat{b}\right),
\ee
where $r_{aa}$ ($r_{bb}$) is the distance from the source to the upper
(lower) slit along beam $a$ ($b$), and $r_{ay}$ ($r_{by}$) is the distance
from the upper (lower) slit to point $y$.
The count rate over all $y$ on the screen is:
\be
\label{Rt}
C(S) \propto \int_S \left\langle E_y^{(-)}E_y^{(+)}\right\rangle dt = 
\int_S \frac{2\epsilon^2}{S}\left(1 + \cos[r_{ay}(t) - r_{by}(t)]\right)dt \sim 
2\epsilon^2,
\ee
assuming that, by the appropriate setting of path length compensator PLC 
in Figure \ref{tri}, $r_{aa} = r_{bb} + nk^{-1}$, for some integer $n$. 
Comparing Eqs. (\ref{Rg}) and (\ref{Rt}), we find $C(S) = C(h)$. Hence, no 
nonlocal effect arises in this case.

The crucial point is to note that the field $E^{(+)}_k$ in Eq. (\ref{fjeld2})
automatically incorporates Feynman's dictum that the modes $a$ and $b$ should
superpose. Thus, we are not free to suppress interference at $k$ in the DA2
set-up in order to forbid the nonlocality. The electric fields in Eqs.
(\ref{fjeld2}) and (\ref{fjeld3}) have been written down in a similar way, by 
including
all modes that are incident at a given point. There is no way, then, to
modify $E^{(+)}_k$ in Eq. (\ref{fjeld2}) without also modifying $E^{(+)}_y$ in Eq. 
(\ref{fjeld3}). Since the latter follows straightforwardly from the second 
quantization formalism \cite{gla63}, and is experimentally confirmed, 
so too are Eq. (\ref{fjeld2}), and the correlation due to it,
inevitable. What remains is to better understand the need for renormalization,
which is taken up in the next section.

\section{Quantum mechanical picture}\label{degenr} 

Although interference experiments rightly belong to the domain of quantum
optics (QO), many of them can usually be translated into quantum mechanical 
language (eg., cf. Ref. \cite{ghz} as regards a number of multiphoton 
interferometric experiments,
and Ref. \cite{dan00,sri154} as regards the delayed choice
experiment).  Sometimes the latter version, even though not entirely valid,
can be easier to physically interpret. To this end, let us revert back to
first quantization formalism and view the optical
modes in the experiment as Hilbert space eigenstates.
Furthermore, for simplicity, we ignore the finite width of the beam, so that
we replace Eq. (\ref{output}) by the simplified version: 
\be
\label{s_output}
|\psi\rangle = \frac{1}{2}\left(|a\rangle + |b\rangle
+ \sqrt{2}|h\rangle\right),
\ee           
where now $|X\rangle$ ($X = a, b, h$) represent the state
corresponding to the photon being found on beam $X$. They constitute the
eigenstates of the 3-dim Hilbert space of the beams.

We then interpret the measurements at DA1 and DA2 as corresponding to two 
different ``which beam?" observables. In this notation,
observable DA1 is given by the spectral decomposition
\be
\label{Og}
\hat{O}_1 = l|a\rangle\langle a| + m|b\rangle\langle b| + p|h\rangle\langle h|,
\ee
where $p$ is the region on the beam $h$ where Bob's detector is placed.
According to the von Neumann projection postulate \cite{vN}, measurement
non-unitarily reduces the state vector to an eigenstate by
the action of one of the projection operators
\be
\label{Ebc}
\hat{P}_a \equiv |a\rangle\langle a|, \hspace{0.5cm}
\hat{P}_b \equiv |b\rangle\langle b|, \hspace{0.5cm}
\hat{P}_c \equiv |h\rangle\langle h|.
\ee
The probability to find the photon at $l$, $m$ or $p$, given by $P(l) = P(m)
= P(h)/2 = 1/4$, is obtained by the expectation value of the corresponding 
projector in the state $|\psi\rangle$ of Eq. (\ref{s_output}). 

Now we can easily check the operational equivalence between the present QM
picture and the quantum optical derivation in Section \ref{main}:
the projectors $\hat{P}_Z$ ($Z = l, m, p$) are equivalent to
the electric field operators $E_X^{(-)}E_X^{(+)}$ ($X = a, b, h$), in the
sense that the $\hat{P}_Z$'s have, respectively, the same
expectation value in the state $|\psi\rangle$ of Eq. (\ref{s_output}), as
the $E_X^{(-)}E_X^{(+)}$'s have 
in the quantum optical state $|\Psi\rangle$ in Eq. (\ref{spdc}), in 
view of Eq. (\ref{Rg}) (apart from a $4\epsilon^2$ factor). 

Measurement in the DA2 basis represents an incomplete 
measurement, because it cannot distinguish between beams $a$ and $b$.
In a non-interferometric setting, the probability of an incomplete 
measurement on $a$ and $b$ is given by the measurement of the projector
\be
\label{lyd} 
|a\rangle\langle a| + |b\rangle\langle b| = \hat{P}_l + \hat{P}_m .  
\ee
If valid, it would indeed suffice to prohibit the nonlocality discussed above.
However, lacking cross-beam terms,
it would also imply that the beams $a$ and $b$ don't interfere 
at $k$, which, as noted earlier, is not in keeping with Feynman's
aforementioned dictum (that indistinguishable transition processes should
interfere), nor consistent with the quantum optical formalism. 

Taking a clue from the form of the electric field operator 
$E_k^{(+)}$ given by Eq. (\ref{fjeld2}), it is straightforward to find that,
to be equivalent to the operator $E_k^{(-)}E_k^{(+)}$ in
Eq. (\ref{fjeld2}), the incomplete projector corresponding to Alice's DA2 
measurement should have the form:
\be
\label{Ea}
\hat{P}_k \equiv \hat{P}_{l+m} = 
(\alpha|a\rangle + \beta|b\rangle)(\alpha^{*}\langle a| + 
\beta^{*}\langle b|) \ne \hat{P}_l + \hat{P}_m, 
\ee
in view of Eq. (\ref{Ebc}). In Eq. (\ref{Ea}), we set the phase factors
$\alpha = e^{ik(r_{ak})}$ and $\beta =
e^{ik(r_{bk})}$ in order that $\langle \hat{P}_k\rangle$ with
respect to state $|\psi\rangle$ should
yield an (unnormalized) probability (= 1) in consonance with Eq. (\ref{one})
and Eq. (\ref{Rf}), apart 
from the $4\epsilon^2$ factor. $\hat{P}_k$ is the required incomplete 
projector that takes into account the
interference between beams $a$ and $b$. It yields the correct expression even
when the two beams are out of phase by $\phi$.

We are now in a position to appreciate why the renormalization is inevitable.
The Hilbert space available to the experiment is 
spanned by $|X\rangle$ ($X = a, b, h$). The projectors
$\{\hat{P}_l, \hat{P}_m, \hat{P}_p\}$ for the DA1 measurement basis are complete
in the sense that 
\be
\label{complete}
\hat{P}_l + \hat{P}_m + \hat{P}_p = I,
\ee
 where $I$ is the
identity operator. On the other hand, the projectors for the DA2 measurement, 
namely $\{\hat{P}_k, \hat{P}_p\}$, clearly do not constitute a complete basis set
since 
\be
\label{incomplete}
\hat{P}_k + \hat{P}_p \ne I.
\ee
They complete a subspace of lower dimensionality (= 2), spanned by vectors 
$\alpha|a\rangle + \beta|b\rangle$ and $|h\rangle$, to which $|\psi\rangle$ is
projected when $\hat{O}_2$ is measured. That the completeness of
basis is necessary to prohibit nonlocal transmission of classical information
was first explicitly noted in Ref. \cite{ghi88}.

One can press ahead with the quantum mechanical picture and, in analogy with
Eq. (\ref{Og}), write down a formal 
spectral decomposition for the DA2 `observable'
\be
\label{Of}
\hat{O}_2 = k\hat{P}_k + p\hat{P}_p.
\ee
The normalization of a state vector is tailored to yield the true probabilities 
only for, and necessarily for, measurements
in a complete basis. This is true irrespective of whether the measurements
are complete or incomplete (i.e, represented by the projectors of the type
Eq. (\ref{lyd})), projective or effect valued. Hence, measurement in the
incomplete basis of $\hat{O}_2$ requires the renormalization. For the same 
reason, $\langle \hat{O}_2 \rangle$ yields an expectation value that must be 
adjusted for normalization (here by a factor of 2/3).
Reverting back to second quantization, we attribute the renormalization to the 
fact that the electric field operators, $E^{+}_k$ and $E^{+}_p$,
in the DA2 set-up do not determine a complete basis in the Fock space
available to the experiment. We note that as an intrinsic limitation of the
first quantization picture, the projectors $\hat{P}_X$ ($X = l, m, p, k$) are to
be used only for calculating probabilities, and not for predicting the state in
which the photon is left if $\hat{P}_X$ is measured, a situation that is
independent of the dynamical and statistical properties of QM.   

No renormalization is needed for the Young double slit measurement
on beams $a$ and $b$, because it does not 
correspond to an incomplete basis. It merely scrambles path information.
The projector for a given point $y$ on the double slit screen in the
`beam' basis is $\hat{P}_k(y) \equiv 
S^{-1}(\alpha(y)|a\rangle + \beta(y)|b\rangle)(\alpha^{*}(y)\langle a| + 
\beta^{*}(y)\langle b|)$. The completeness of the basis set follows from 
seeing that
\be
\int_S 
S^{-1}(\alpha(y)|a\rangle + \beta(y)|b\rangle)(\alpha^{*}(y)\langle a| + 
\beta^{*}(y)\langle b|)dy = \hat{P}_l + \hat{P}_m,
\ee
where $\alpha(y)$ and $\beta(y)$ are the phase of the modes $a$ and $b$ at
point $y$. Let us disambiguate the twin usages of the word completeness in the 
present context: 
$\hat{O}_2$ corresponds to an {\em incomplete measurement} in an
{\em incomplete basis}. It is of course the latter feature that is responsible
for the nonlocality. On the other hand, $\hat{O}_1$ corresponds to an incomplete
measurement in a complete basis. 
 
By the action of $\hat{P}_k$, 
the state vector is said to undergo `coherent projection' or
`coherent reduction' into a Hilbert space of lower dimensionality. In
contrast, no such reduction occurs for `projective reduction', obtained by the 
action of $\hat{P}_l + \hat{P}_m$. Coherent reduction is characteristic of 
{\em controlled incomplete} interferometric measurements. It brings further 
richness to quantum measurement and the ways that it can be used to
exploit the essentially quantum phenomenon of superposition. What is 
encouraging is 
that quantum optical tests for it are well feasible, and can indeed be already
gleaned from existing interferometric experiments.
 
That Alice may transmit a nonlocal classical signal by choosing to measure
in a complete or incomplete basis (in general, bases incomplete in different
eigenstates) is not incompatible with the no-signaling arguments of 
Refs. \cite{ghi88,ebe78,ghi80,ebe89,boh93,bus82,jor83,gar90,can78}, 
because, as pointed out by Ghirardi et al. \cite{ghi88}, the assumption of 
measurement in a complete basis is implicit in them. Moreover, they deal with 
entangled multipartite rather than single particle systems. In view of the
transparent arguments leading to, but surprising conclusions following from, 
the general measurement reported here, its experimental test
is quite valuable. As an indirect evidence for its feasibility, an
entanglement-based version of this effect was reported in
Ref. \cite{sri148}, wherein it is argued that the results reported by
Zeilinger \cite{zei00} correspond to measurement in incomplete bases. 
However, we note that from the viewpoint of implementation and practical
application the single particle version is much simpler.

\section{Conclusion}

Feynman noted that the central mystery of QM--
namely, superposition-- is encapsulated by the double-slit interference 
\cite{fey65}. Coherent reduction, as discussed above,
adds a further perspective to this `mystery'. 
If confirmed, it makes for an interesting addition
to the basic quantum communication acts: the nonlocal transmission of a 
bit without concomitant communication, classical or quantum, along a causal
channel. 


\vspace*{2.0cm}

\end{document}